\documentclass[aps,pre,showpacs,reprint]{revtex4-1}
\usepackage[]{graphicx}
\usepackage{color}

\begin{document}

\title{Fundamental issues in nonlinear wideband-vibration energy harvesting}

\author{Einar Halvorsen}
\email[E-mail: ]{Einar.Halvorsen@hive.no}
\affiliation{Department of Micro and Nano Systems Technology, Faculty of Technology and Maritime Sciences,  
Vestfold University College, P.O. Box 2243,  N-3103 T\o nsberg, Norway}

\date{\today}

\begin{abstract}
Mechanically nonlinear energy harvesters driven by broadband vibrations 
modeled as white noise are investigated. We derive an upper bound on output 
power versus load resistance and show that, subject to mild 
restrictions that we make precise, the upper-bound performance can be 
obtained by a linear harvester with appropriate stiffness. 
Despite this, nonlinear harvesters can have implementation-related advantages. 
Based on the Kramers equation, we numerically obtain the output power at 
weak coupling for a selection of phenomenological elastic potentials 
and discuss their merits.
\end{abstract}


\maketitle

\section{Introduction}
Energy harvesting from motion is a means to power wireless 
sensor nodes in constructions, machinery and on the human body 
\cite{Beeby2006,Mitcheson2008}. A vibration energy harvester contains a 
proof mass whose relative motion with respect to a frame drives a transducer that generates electrical power. 
Linear resonant devices are superior when driven by harmonic vibrations 
at their resonant frequency, but perform poorly for off-resonance conditions. 
As real vibrations may display a rich spectral content,
 sometimes of broadband nature,
 there has been considerable interest in using nonlinear suspensions 
to shape the spectrum of the harvester's response to better suit the vibrations \cite{Burrow2007,Cottone2009,Gammaitoni2009,Erturk2009c,Soliman2008,Stanton2009,Marzencki2009,Marinkovic2009,Nguyen2010,Nguyen2011}. The wider spectral response of nonlinear devices is expected to be  beneficial for broadband vibrations.

The studies so far indicate some advantages of nonlinearities for broad-banded vibrations, but little is known about which conditions make a nonlinear harvester favorable compared to a linear one. This is due to lack of adequate theory and due to the studies being concerned about specific experimental or numerical examples of nonlinear harvesters that are compared to specific examples of linear harvesters that could have been chosen differently. Furthermore, several studies do not consider the role of electrical loading which is known to have a dramatic influence on the consequences of mechanical nonlinearities for the output power \cite{Halvorsen2008}.

White noise is widely used in physics and engineering 
\cite{Kampen2007,CloughPenzien1993,Haykin1983,Aastrom1970}, and 
is also important in studying  broadband energy harvesting 
\cite{Lefeuvre2006,Halvorsen2008,Scruggs2009,
Tvedt2010,Daqaq2010,Ali2011,Daqaq2012}. 
If the vibration spectrum is flat 
over the frequency range of the
harvester, the harvester itself provides a cut-off  
making the infinite bandwidth of white noise a meaningful 
idealization. White noise approximates colored noise with correlation 
time sufficiently short compared to the characteristic times of the 
system. Aspects of a nonlinear harvester's 
performance hinging on a finite correlation time and not present 
for white noise are, albeit interesting, 
necessarily relying on a limited 
vibration bandwidth. Therefore white noise 
is a good case for investigating broadband performance.

Here we investigate theoretically the behavior of mechanically nonlinear 
energy harvesters driven by a Gaussian white noise  acceleration.
We derive rigorous upper bounds on the output power for arbitrary elastic potential and show that subject to mild restrictions on the device parameters, it is possible to find a linear device that performs equally well as the upper bound.  
We give a compact expression for the output power that we use to numerically investigate the weak coupling limit of harvesters for different quartic polynomial potentials taking electrical loading fully into account.      

\section{Model and notation}

\begin{figure}[t]
\includegraphics[]{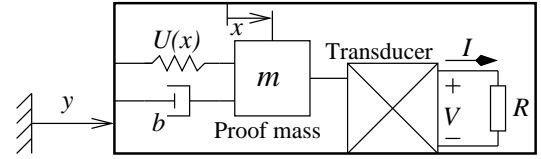}
\caption{Vibration energy harvester model.}
\label{fig:ehsystem}
\end{figure}

An energy harvester model that isn't technology specific is shown in Fig. \ref{fig:ehsystem}. The corresponding state space equations with a linear 
electromechanical transducer and a nonlinear mechanical suspension can 
be written
\begin{eqnarray}
\dot x & = & v \label{eq:sp1}\\
m\dot v & =&  -U'(x) -\Gamma q/C - bv + m a, \label{eq:sp2} \\
 -R\dot q & = & RI=V= \Gamma x/C + q/C, \label{eq:ohm} \label{eq:sp3}
\end{eqnarray}
where $m$ is the proof mass, $x$ its relative displacement, $v$ its velocity, 
$U$ the open-circuit internal 
energy, $q$  the transducer-electrode charge, 
$V$ the output voltage, $I$ the current,
$b$ the damping coefficient,
$R$ the load resistance, $C$ the clamped capacitance and $\Gamma$ the transduction factor.
The device-frame acceleration $\ddot y=-a$ is Gaussian white noise with a two-sided spectral density $S_a$.
The equations can represent
a piezoelectric or an electrostatic energy harvester. An electromagnetic 
harvester gives the same mathematical structure, but different physical interpretation.  We use charge as the independent variable \cite{Tilmans1996}. 
Using voltage instead is physically equivalent and also common, see e.g. \cite{Guyomar2005}.  

Ensemble averages with respect to the stationary distribution generated 
by the process (\ref{eq:sp1}-\ref{eq:sp3}) will be 
denoted by $\langle\cdots\rangle$. 
The mean output power 
$P=\langle V^2 \rangle/R$ will be our main object of interest. 
A number of other expressions for $P$ immediately follow 
 by using stationarity, (\ref{eq:sp1}) and 
(\ref{eq:sp3}). We will use some of these expressions without giving the 
derivation. All results for linear systems are exact and taken 
from \cite{Halvorsen2008} unless said otherwise.
 
From (\ref{eq:sp3}), 
$q={\cal O}(\Gamma)$. The second term on the right hand side of (\ref{eq:sp2}) is ${\cal O}(\Gamma^2)$
and can then be dropped 
in the limit $\Gamma\rightarrow 0$. This is the weak coupling limit, 
which in the stationary state has the reduced probability density 
\begin{equation}
W_\mathrm{st}^0(x,v)=\exp(-bv^2/mS_a-2bU(x)/m^2S_a)/Z_\mathrm{st}^0
\label{eq:wst0}
\end{equation} 
where $Z_\mathrm{st}^0$ is a normalization constant \cite{Gardiner2004}. 
We denote expectations in this limit by $\langle \cdots \rangle_0$.

\section{Bounds and limits}
In this section, we  prove that a previously known lemma on the mechanical input power of linear harvesters also encompasses  mechanically nonlinear ones, and discuss its consequences.  We then show that known asymptotic formulas for large or small load resistances are upper bounds on output power. Finally we find improved bounds that are asymptotically correct in both limits and compare to exact results for a linear harvester. 

\subsection{Power balance}
\label{sec:powerbalance}
The important observation that the mean input power is  
$P_\mathrm{in}=mS_a/2$ was made in \cite{Scruggs2009} where it was proved 
for linear harvesters.  
For our nonlinear system and  $\Gamma=0$, 
all power is dissipated in the damper, (\ref{eq:wst0}) implies the equipartition theorem,   
and $P_\mathrm{in}=b\langle v^2\rangle_0=mS_a/2$.
For general $\Gamma$,
consider the input energy $\int_{t_1}^{t_2}mav\mathrm{d}t$ over a time interval.  
When the actually continuously differentiable $a$ is modeled as white noise, the appropriate stochastic representation of the energy is a Stratonovich integral $mv\circ\mathrm{d}a$ \cite{Oeksendal2007}.  
We have $mv\circ\mathrm{d}a=mv\mathrm{d}a+mS_a\mathrm{d}t/2$ where $v\mathrm{d}a$ is an Ito integral and has zero expectation \cite{Gardiner2004,Ko2006}. The input-energy expectation is then $mS_a\mathrm{d}t/2$ which yields the stated expression for $P_\mathrm{in}$. 

The observation means that $\eta=2P/mS_a$ is an efficiency that should be maximized, as opposed to linear narrow-band harvesting where power transfer is maximized.
It also implies a power balance 
\begin{equation}
P=mS_a/2-b\langle v^2\rangle.
\end{equation}

For linear harvesters, 
 $\eta\rightarrow 1$ as $k^2Q_\mathrm{m}\rightarrow\infty$ 
where $k^2=\Gamma^2/KC \le 1$ is the transducer 
electromechanical coupling factor, $K$ is the open-circuit stiffness and $Q_\mathrm{m}$ is the open-circuit 
quality factor\cite{Halvorsen2008}. Hence, it is impossible to improve 
significantly on a linear harvester 
that is already very efficient.
The device in \cite{Goldschmidtboeing2011} for example,
has $k^2Q_\mathrm{m}\approx 7.8$ resulting in $\eta \approx 0.79$. 
The great number of harvesters, especially those with small volume,
that perform substantially 
below their theoretical maximum \cite{Mitcheson2008}, suggests that the weak coupling regime nevertheless has great practical relevance. 

\subsection{Asymptotic formulas as bounds}
The load resistance determines the electrical time scale $\tau=RC$ 
distinguishing different regimes of operation.
When $\tau$ is the fastest scale, i.e. $\tau \rightarrow 0$, we have \cite{Halvorsen2008} 
\begin{equation}
P \sim  \Gamma^2\langle v^2\rangle \tau/C  
  \sim  \Gamma^2\langle v^2\rangle_0 \tau/C 
   = \Gamma^2\tau mS_a/2bC. 
\label{eq:powsmalltau}
\end{equation}
 From (\ref{eq:sp3}), it is readily proved that 
$P=\Gamma^2\tau\langle v^2\rangle/C-\tau^3\langle \dot I^2\rangle/C
\le\Gamma^2\tau\langle v^2\rangle/C$. 
One can also show that $\langle v^2\rangle\le\langle v^2\rangle_0$.
Hence, both asymptotic relations in 
 (\ref{eq:powsmalltau}) are upper bounds on the output power. 
We note that the bounds are valid for any $U$ that permits a stationary 
distribution and that the output power is otherwise independent of $U$ 
when $\tau\rightarrow 0$.

When the electrical time scale is the slowest in the system, i.e. 
when $\tau\rightarrow\infty$, we have \cite{Halvorsen2008,Gammaitoni2009} 
\begin{equation}
P \sim \Gamma^2\langle (x-\langle x \rangle)^2 \rangle /\tau C 
\sim \Gamma^2\langle (x-\langle x \rangle_0)^2 \rangle_0 /\tau C. 
\label{eq:powlargetau}
\end{equation}  
The leftmost asymptotic formula in (\ref{eq:powlargetau}) is also an upper bound. 
This is seen by using (\ref{eq:sp3}) to find 
\begin{equation}
P=\Gamma^2\langle (x-\langle x \rangle)^2 \rangle /\tau C -\langle (q-\langle q \rangle)^2 \rangle /\tau C 
\label{eq:powxq}
\end{equation}
which gives the inequality when dropping the second term. The 
rightmost asymptotic formula in (\ref{eq:powlargetau}) need not be an upper bound as can be inferred already from linear theory. 
We note that (\ref{eq:powlargetau}), in contrast to (\ref{eq:powsmalltau}),  is strongly dependent on $U$ as it is proportional to $\langle (x-\langle x \rangle)^2 \rangle $.

The maximum power as 
a function of $\tau$ must necessarily be found at an intermediate value of $\tau$ between the small-$\tau$ and large-$\tau$ regimes. Since the output power 
is respectively insensitive and sensitive to the nature of $U$ in these two regimes, the degree to which the maximum power can be improved by mechanical nonlinearities is an open question. 

\subsection{Improved power bounds and the linear case}
We now address the potential benefits of nonlinear devices by deriving improved power bounds and comparing to linear behavior. 
Define 
$z=q-\langle q \rangle  -D(x-\langle x \rangle) -Bv$ and find the values of the constants $B$ and $D$ that minimize $\langle z^2 \rangle$. Eliminate covariances between 
$x$ and $q$ using $\Gamma\langle xq \rangle +  \langle q^2 \rangle = 0 $ 
and use $P=\Gamma\langle qv\rangle/C$ and (\ref{eq:powxq}) to write the minimum value as
\begin{equation}
\langle z^2 \rangle =  
\frac{\tau C}{\Gamma^2}\frac{\langle(q-\langle q \rangle)^2\rangle}{\langle (x-\langle x \rangle)^2 \rangle}P -\frac{C^2P^2}{\Gamma^2\langle v^2 \rangle}.
\end{equation}
Next, use this to eliminate the variance of $q$  in (\ref{eq:powxq}) and rearrange to obtain 
$P = (1-\langle z^2 \rangle/\tau C P)P_\mathrm{u1}\le P_\mathrm{u1}$
where 
\begin{equation}
P_\mathrm{u1} = 
 \frac{\Gamma^2}{C}
  \frac{\tau \langle v^2 \rangle  \langle (x-\langle x \rangle)^2 \rangle}
  {\langle (x-\langle x \rangle)^2 \rangle+\tau^2\langle v^2 \rangle}.
\label{eq:ubound}
\end{equation}
We see that (\ref{eq:ubound}) agrees with (\ref{eq:powsmalltau}) and (\ref{eq:powlargetau}) in their respective limits and is a tighter bound. 

The quantity $\omega_\mathrm{m}=\sqrt{\langle v^2 \rangle/ \langle (x-\langle x \rangle)^2 \rangle}$ can be used to eliminate the displacement variance in (\ref{eq:ubound}). 
Using $P=mS_a/2-b\langle v^2 \rangle\le P_\mathrm{u1}$ we find a \textit{lower} bound on $\langle v^2 \rangle$ which we 
substitute back into the power balance equation to obtain $P\le P_\mathrm{u2}$ where the new bound is  
\begin{equation}
P_\mathrm{u2} = \frac{mS_a}{2}\frac{\Gamma^2\tau/Cb}{1+\Gamma^2\tau/Cb+\omega_\mathrm{m}^2\tau^2}.
\label{eq:ubound2}
\end{equation}
$P_\mathrm{u2}$ is manifestly less than $P_\mathrm{in}$ and is asymptotically approaching the exact result at both the extreme limits of $\tau$. 
 
We can interpret $\omega_\mathrm{m}$ as the root-mean-square frequency  of the spectrum \cite{Barnes1993} of the displacement $\delta x=x-\langle x\rangle$. This follows from representing the variances in terms of the spectral densities $S_{\delta x\,\delta x}(\omega)$ and $S_{vv}(\omega)=\omega^2S_{\delta x\,\delta x}(\omega)$ of $\delta x$ and $v$ respectively, 
that is 
\begin{equation} 
\omega_\mathrm{m}^2 =\frac{\langle v^2\rangle}{\langle\delta x^2\rangle}=\frac{\int_{-\infty}^\infty\omega^2S_{\delta x\,\delta x}(\omega)\mathrm{d}\omega/2\pi }{\int_{-\infty}^\infty S_{\delta x\,\delta x}(\omega)\mathrm{d}\omega/2\pi }.
\end{equation}  

The most optimistic estimate of output power permitted by ($\ref{eq:ubound2}$) 
is found for load resistances such that $\omega_\mathrm{m}\tau=1$ and is 
\begin{equation}
P_\mathrm{u2,Opt}=(mS_a\Gamma^2/2Cb)/(2\omega_\mathrm{m} + \Gamma^2/Cb).
\label{eq:Pu2Opt} 
\end{equation}
This can be compared to the exact  output 
power of an optimally loaded linear harvester which  is 
\begin{equation}
P_\mathrm{Lin,Opt}=(mS_a\Gamma^2/2Cb)/(2\omega_0 + b/m + \Gamma^2/Cb)
\label{eq:PLinOpt}
\end{equation}
where $\omega_0$ is the open-circuit resonance. 
The two power expressions differ only in terms in the denominators: $2\omega_\mathrm{m}$ in (\ref{eq:Pu2Opt}) v.  $2\omega_0+b/m$ in (\ref{eq:PLinOpt}). 
 With all other parameters except load resistance held equal, a linear system can therefore be made to perform better than, worse than or equally to the bound depending on its stiffness. It will meet the performance 
of the bound if its stiffness is such that  $\omega_0=\omega_{\mathrm{m}}-b/2m$. 
 The only fundamental restriction on the linear system is that it is stable, i.e. has $k^2<1$ \cite{Tilmans1996} which is equivalent to $\omega_0^2\ge\Gamma^2/mC$.
Hence, a linear device meeting the bound is
realizable 
if 
\begin{equation}
\omega_\mathrm{m}>b/2m+|\Gamma|/\sqrt{mC}.
\label{eq:lincrit}
\end{equation}
\textit{Therefore nonlinear 
harvesters are not fundamentally better than linear ones.}

Harvesters that have their spectrum shaped by nonlinear design of their proof mass suspension will, like linear resonant devices, typically be designed to have  $b/2m$ much less than the characteristic frequencies of proof-mass motion in order to maximize performance. We therefore expect $b/2m\ll\omega_\mathrm{m}$ to be a typical case for such nonlinear devices.  A corresponding linear system performing equally to the bound, will then have $\omega_0\approx\omega_\mathrm{m}$. That is, its resonance lies within the frequency range of the nonlinear harvester's spectrum. 

We note that failure to fulfill the criterion (\ref{eq:lincrit}) because of the second term on the r.h.s, corresponds to coupling strong enough that a linear device is not an alternative due to lack of stability or due to being only marginally stable.
We would expect this situation for truly nonresonant devices with low damping. 
For $\Gamma$ approaching this limit from below, one has the high-efficiency situation  discussed in section \ref{sec:powerbalance} even with considerable damping (moderate $Q_\mathrm{m}$ for the linear device). 

While (\ref{eq:ubound2}) is always an upper bound on the output power, it is quite possible that this bound is a poor approximation and considerably overestimates the actual output power. We might expect this situation when the spectrum $S_{\delta x\, \delta x}$ has multiple peaks widely separated in frequency such as for quartic bistable potentials \cite{Voigtlaender1985,Dykman1988}. If so, the actual performance can be met by a linear device with $\omega_0$ larger than $\omega_\mathrm{m}-b/2m$ by an amount in correspondance to the degree of overestimate. This has to be checked for each particular case. The criterion (\ref{eq:lincrit}) is a sufficient, but not necessary, condition for the realizability of a linear harvester that performs equally well or better than a harvester characterized by $\omega_\mathrm{m}$.

\section{Numerical results}
We now consider how to directly calculate the output power for concrete examples. 
From 
(\ref{eq:sp1}) and (\ref{eq:sp3}) it follows that 
$V=(\Gamma/C)\int_{-\infty}^t\exp(-(t-t_1)/\tau)v(t_1)\mathrm{d}t_1$. Inserting this expression into $P=\Gamma\langle v(t)V(t)\rangle$, we obtain 
\begin{equation} 
P=  \frac{\Gamma^2}{C}\int_0^\infty e^{-t/\tau}\langle v(t)v(0) \rangle \mathrm{d}t =
\frac{\Gamma^2}{C}\tilde K_{vv}(1/\tau),
\end{equation}
i.e. that the output power is proportional to the Laplace transform $\tilde K_{vv}$ 
of the velocity autocorrelation 
function.

In the weak coupling limit $\Gamma\rightarrow 0$,
we can approximate $\tilde K_{vv}$ by its value  $\tilde K_{vv}^0$  
for $\Gamma=0$ to obtain the leading order.  $\tilde K_{vv}^0$  
can be found from the transition probability by solving the Fokker-Planck equation corresponding to
(\ref{eq:sp1}) and (\ref{eq:sp2}) with $\Gamma=0$, i.e. the Kramers equation \cite{Risken1996}. 

Without pursuing it further, we remark that an alternative method to calculate 
the output power, and therefore also  $\tilde K_{vv}^0$, would  be to find 
a \textit{stationary} solution of the Fokker-Planck equation for the energy 
harvester\cite{Halvorsen2008} in the weak coupling limit 
and use $P\sim\Gamma\langle qv\rangle_0/C$ or 
$P\sim\Gamma\langle vV\rangle_0$. 

\subsection{Numerical method}
We determine $\tilde K_{vv}^0$ numerically from 
the Kramers equation by orthogonal function 
expansions and matrix continued fraction methods following 
\cite{Voigtlaender1985,Risken1996}. The spatial basis functions are 
$\psi_n(x)=\sqrt{W(x)}\pi_n(x),\; n=0,1,...$ where $W(x)=\exp(-2bU(x)/m^2S_a)/Z_0$, $Z_0$ is a normalization constant and $\pi_n(x)$ are orthonormal polynomials with $W(x)$ as weight function. We express all spatial-basis matrix elements in terms of the recurrence coefficients for $\pi_n$
which are determined by adapting the Lanczos method described in \cite{Gautschi2005}
to continuous variables. 
 Dimensionless variables distinguished by asterisk subscripts and based on a characteristic length scale $l_\mathrm{s}$ and frequency scale $\omega_\mathrm{s}$ are used, e.g. $P_*=P/ml_\mathrm{s}^2\omega_\mathrm{s}^3$, $\Gamma_*=\Gamma/\sqrt{m\omega_\mathrm{s}^2C}$, $S_{a*}=S_a/l_\mathrm{s}^2\omega_\mathrm{s}^3$ and $\tau_*=\omega_\mathrm{s}\tau$.

\subsection{Symmetric quartic potentials}
\begin{figure}[t]
\includegraphics[]{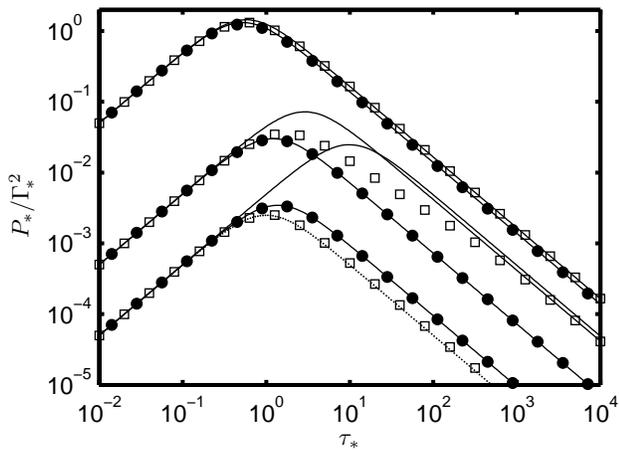}
\caption{Output power $P_*$ versus electrical time scale $\tau_*$ for mono- and bistable potentials at weak coupling, $\gamma_*=0.01$, $S_{a*}=10^{-4}, 10^{-3}, 0.1$ (bottom to top) and $B_*=1$. Open squares: Numerical solution for $A_*=-0.5$. Solid circles: Numerical solution for $A_*=0.5$. Solid lines: Corresponding upper bounds. Dotted line: Solution from linearization around potential minimum with stiffness $2|A_*|$.}.
\label{fig:pvr}
\end{figure}
\begin{figure}
\includegraphics[]{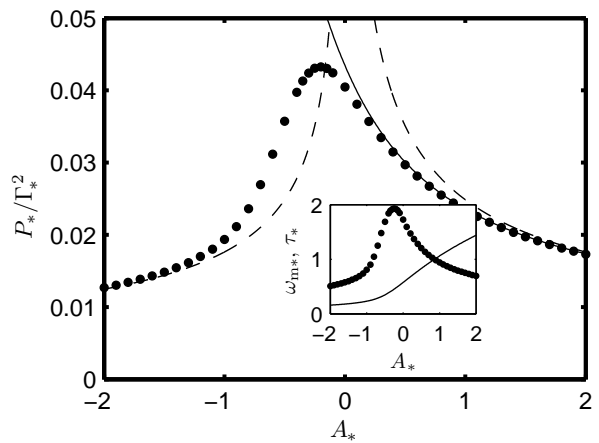}
\caption{Maximum output power $P_{*}$  as a function of $A_*$ at weak coupling, $\gamma_*=0.01$, $B_*=1$  and $S_{a*}=10^{-3}$. 
Solid circles: Numerical solution. Dashed lines: Solution from linearization around potential minima with stiffness $|2A_*|$ or $|A_*|$. Thin solid line: upper bound.
Inset shows corresponding optimal load given by $\tau_*$ (solid circles) and the root-mean-square frequency $\omega_{\mathrm{m}*}$ (solid line).
}
\label{fig:optpow}
\end{figure}
We first consider the much studied symmetric quartic potential $U=Ax^2/2+Bx^4/4$, choose $l_\mathrm{s}$ such that $B_*=Bl_\mathrm{s}^2/m\omega_\mathrm{s}^2=1$ and $\omega_\mathrm{s}$ such that $\gamma_*=b/m\omega_\mathrm{s}=1/100$.   
Traces for a bistable potential with $A_*=A/m\omega_\mathrm{s}^2=-0.5$ and a monostable potential 
with $A_*=0.5$ are shown in Fig. \ref{fig:pvr}.
For small values of $\tau_*$, the output power 
collapses as predicted by (\ref{eq:powsmalltau}) onto the same asymptotic form for both potentials. 
For $S_{a*}=1.0\cdot 10^{-4}$ the mass vibrates around a potential minimum, 
 giving a performance for larger $\tau_*$ that differs between the two cases due to their different linear stiffnesses at the minima, i.e. $2|A_*|$ for the bistable potential and $A_*$ for the monostable potential. At $S_{a*}=0.1$,
the quartic term in the potential determines the behavior. In the intermediate case $S_{a*}=1.0\cdot 10^{-3}$, the two potentials give comparable maximum power even though there is a considerable difference between them for large $\tau_*$. 

For weak coupling, the upper bounds (\ref{eq:ubound},\ref{eq:ubound2}) simplify to 
\begin{equation}
P_\mathrm{u1}=P_\mathrm{u2}=(mS_a\Gamma^2\tau/2Cb)/(1+\omega_\mathrm{m}^2\tau^2)
\label{eq:uboundwc}
\end{equation}
with  $\omega_\mathrm{m}^2=\langle v^2 \rangle_0/ \langle (x-\langle x \rangle)^2 \rangle_0$. In this limit we can calculate $\omega_\mathrm{m}$ directly from the known expression for $\langle v^2\rangle_0$ and the value of $\langle (x-\langle x\rangle_0)^2\rangle_0$ obtained from numerical quadrature using (\ref{eq:wst0})
as the probability density. Then  $\omega_\mathrm{m}$ is independent of $\tau$, but does depend on $S_a$. We have $\langle U'(x)(x-\langle x\rangle_0)\rangle_0=m\langle v^2\rangle_0$ so $m\omega_\mathrm{m}^2$  corresponds to the stiffness in standard stochastic equivalent linearization \cite{Crandall2001}. 
The bound has a maximum value of $mS_a\Gamma^2/4Cb\omega_\mathrm{m}$  at $\tau=1/\omega_\mathrm{m}$. The maximum value will therefore increase and shift to a larger $\tau$ when $\omega_\mathrm{m}$ is lowered. As $\omega_\mathrm{m}$ can be strongly dependent on the acceleration spectral density $S_a$, the bound can have a nontrivial dependence on $S_a$. For example, for $S_{a*}=10^{-4}$ and $S_{a*}=10^{-3}$ in Fig. \ref{fig:pvr}, we find respectively $\omega_\mathrm{m*}=\omega_\mathrm{m}/\omega_\mathrm{s}=0.101$ and $\omega_\mathrm{m*}=0.347$ for the bistable potential. This frequency difference is big enough for the bounds to cross. 

The value $S_{a*}=10^{-4}$ is small enough that the proof mass exhibits approximately linear dynamics around the potential minima, as indicated by the agreement between the dotted line in the figure and the numerical calculation. 
The
root-mean-square displacement is then on the order of the half the separation between the potential minima for the bistable system,  $\omega_\mathrm{m}^2\approx mS_aB/2b|A|$,  $m\omega_\mathrm{m}^2$ is very different from the linear stiffness $2|A|$, 
and the bound grossly overestimates the actual performance.
At small $S_{a*}$,
the longest time scale is that of interwell transitions 
as given by Kramers' rate problem \cite{Kampen2007,Gammaitoni2009} and 
the large-$\tau$ asymptotics is only reached for $\tau$ values far above the optimum.
This demonstrates the necessity of the more complicated numerical treatment
in predicting maximum power as opposed to bounding it. 

Fig. \ref{fig:optpow} shows the output power versus the parameter $A_*$ when the load is optimized for every $A_*$. The value of the optimal $\tau_*$ in the inset varies correspondingly. Together with the numerical solution and the value of the bound, we show the output for linear devices with stiffness $2|A_*|$ or $A_*$ as an indication of 
when the proof mass mostly vibrates around the potential minima. The 
 values of $\omega_\mathrm{m}$ used to calculate the bound are shown in the inset.
The maximum power is obtained for a negative value of $A_*$, i.e with a bi-stable potential, like demonstrated for a fixed load and colored noise in \cite{Cottone2009}.
But, as the bound 
corresponds to a linear device with $\omega_0=\omega_\mathrm{m}-b/2m$, more power can be obtained with a linear device. 
Increasingly negative $A_*$ again leads to vibrations around the minima with rare interwell transitions as discussed above for small $S_a$, and the bound's overestimate becomes large (leaving the plot). 
For sufficiently negative $A_*$, a linear system with stiffness $2|A_*|$ gives less power. From the monotonic frequency-behavior of (\ref{eq:PLinOpt}), we can then conclude that a linear device with $\omega_0$ somewhat less than $\sqrt{2|A|/m}$, but still larger than $\omega_\mathrm{m}-b/2m$ can match or outperform the bistable harvester.              

For small negative and all positive values of $A_*$ in Fig. \ref{fig:optpow}, linear devices with the same stiffness $A_*$ or $2|A_*|$ as the nonlinear devices have at their potential minima give more power. This can by understood from the quartic term of the potential limiting proof mass motion. We also note that the bound is a good approximation for positive $A_*$, as was also the case in Fig. \ref{fig:pvr}.
  
These considerations show that the motivation for utilizing nonlinear stiffness 
is rather one of necessity than one of advantage. Implementation constraints such as, e.g.,  package size and/or beam dimensioning may prohibit linear operation. In this respect, we can think of the quartic term of the potential as a model of proof mass confinement or beam stretching at large amplitudes.    

\begin{figure}[t]
\includegraphics[]{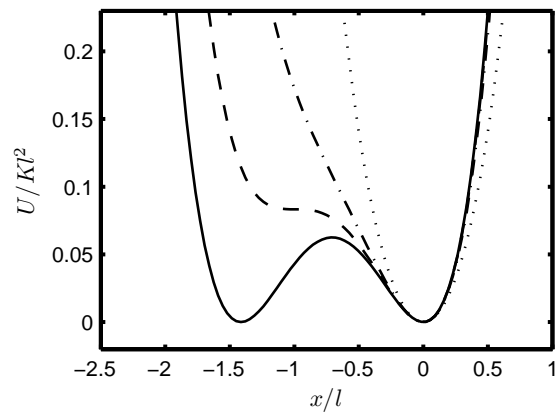}
\caption{Quartic potentials.
Dotted line: $\xi=0$, hardening Duffing spring; Dashed-dotted line: $\xi=\sqrt{2/3}$, negative tangential stiffness arises; Dashed line: $\xi=2\sqrt{2}/3$, bi-stability arises; $\xi=1$ symmetric bistable potential.}
\label{fig:quarticasym}
\end{figure}

\subsection{Asymmetric quartic potentials}
We now consider a suspension made of a stable elastic material without built-in stress, choose $U(0)=0$ and require $U'(0)=0$, $U''(0)>0$ and $U(x)>0\;\forall x\ne 0$. 
The lowest order nontrivial polynomial form 
can then be parametrized as 
\begin{equation} 
U(x)=\frac{1}{2}Kx^2 + \frac{K\xi}{\sqrt{2}l}x^3 + \frac{K}{4l^2}x^4 
\label{eq:quarticasym}
\end{equation}
where $|\xi|<1$, $K>0$  and  $l$ is a length scale,
see Fig. \ref{fig:quarticasym} which illustrates how the potential varies with $\xi$. 
We choose $\omega_\mathrm{s}=\sqrt{K/m}$ and $l_\mathrm{s}=l$ as characteristic scales. 
A linear system with stiffness constrained to the same value $K$ as in (\ref{eq:quarticasym}), and therefore with $\omega_0=\omega_\mathrm{s}$, is used in some comparisons.

\begin{figure}[t]
\includegraphics[]{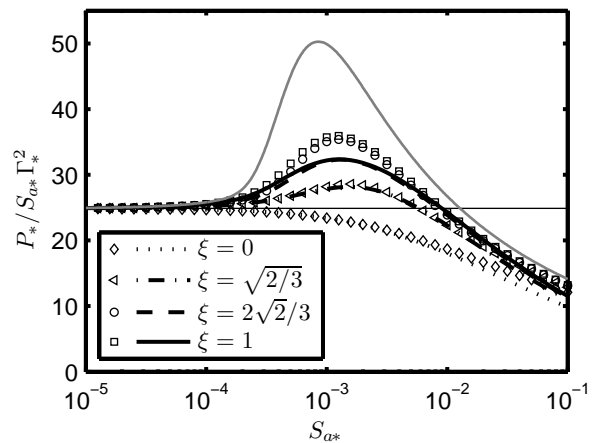}
\caption{Output power $P_{*}$ relative to acceleration spectral density $S_{a*}$ versus $S_{a*}$ both for $\tau_*=1$ (thick lines) and for optimal $\tau_*$ at each point (markers). 
Thin solid line: Linear device with $\omega_0=\omega_\mathrm{s}$. All devices have the same linear stiffness. $\gamma_*=0.01$. Medium thick, grey solid line: upper bound for $\xi=2\sqrt{2}/3$.}
\label{fig:pvs}
\end{figure}

\begin{figure}
\includegraphics[]{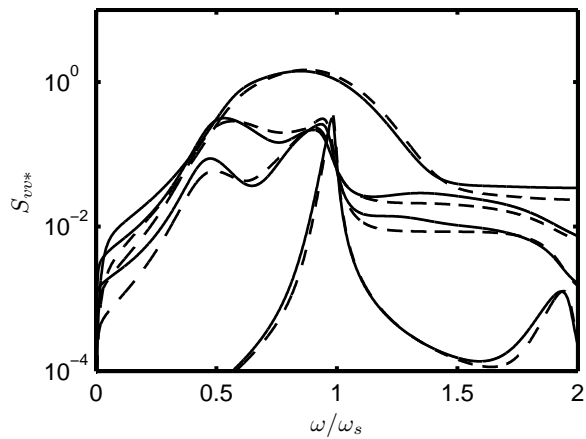}
\caption{Velocity spectral density versus frequency at weak coupling for $S_{a*}=1\cdot10^{-4},5\cdot10^{-4},1\cdot10^{-3},5\cdot10^{-3}$ (from bottom to top traces at the highest frequencies). Solid lines: $\xi=1$. Dashed lines: $\xi=2\sqrt{2}/3$.  
}
\label{fig:spect}
\end{figure}

Figure \ref{fig:pvs} compares output power as  function of acceleration spectral density $S_{a*}$ for harvesters with different values of the parameter $\xi$. To ease comparison the power is divided by $S_{a*}$. A linear harvester then appears as a horizontal line as shown for the particular case with $\omega_0=\omega_\mathrm{s}$.
For each nonlinear potential, results are shown both with fixed load $\tau_*=1$  (lines) and with $\tau_*$ optimized at each value of $S_{a*}$ (markers). $\tau_*=1$ is optimal for the linear system with $\omega_0=\omega_\mathrm{s}$, and therefore for all the shown potentials at small $S_{a*}$. The difference in output power between the two loading cases are moderate for these examples. It is  largest for the largest values of $\xi$ which have the lowest $\omega_\mathrm{m}$. For example, for $S_{a*}=10^{-3}$ we have $\omega_{\mathrm{m}*}=1.061,0.793,0.496$ and $0.347$ from lowest to highest $\xi$. From these values we also note that increased power correlates with lower $\omega_\mathrm{m}$ as we would expect from the form of the bound (\ref{eq:uboundwc}).
 
Fig.  \ref{fig:pvs} shows that
the nonlinear devices with $\xi\ne 0$ give an $S_{a*}$-range of better performance than their linear counterpart with $\omega_0=\omega_\mathrm{s}$. This is the case even with $\tau_*=1$ which is optimal only for that linear device. The consistently lower power for $\xi=0$ is due to the stiffening nature of the potential which limits motion and shifts the spectrum to higher frequencies. The other potentials have a range of softening behavior causing a shift to lower frequencies and higher power.    

Also shown on dimensionless form in Fig. \ref{fig:pvs} (grey line), is (\ref{eq:uboundwc}) for $\xi=2\sqrt{2}/3$ evaluated with $\tau=1/\omega_\mathrm{m}$. Each point of this curve represents an optimally loaded linear device with open-circuit frequency $\omega_0=\omega_\mathrm{m}-b/2m$. For  $S_{a*}=10^{-3}$, this corresponds to $\omega_0=0.496\omega_\mathrm{s}-0.005\omega_\mathrm{s}
\approx 0.5\omega_\mathrm{s}$. If we compare to a linear system with $\omega_0=0.5\omega_\mathrm{s}$ instead of one with $\omega_0=\omega_\mathrm{s}$, it has $P_*/S_{a*}\Gamma_*^2\approx 50$ outperforming all nonlinear cases in Fig. \ref{fig:pvs} over all values of base acceleration spectral density $S_{a*}$. 

The comparison between nonlinear and linear suspensions to judge their relative merits is only fair if the harvester responses are within approximately the same frequency range. In the preceding analysis we secured that by choosing the open-circuit frequency of the linear device approximately equal to the $\omega_\mathrm{m}$ of the nonlinear device. We also discussed how this condition could be relaxed for weakly excited bistable systems. To be more specific on the spectral characteristics, the velocity spectral density $S_{vv}(\omega)=2\mathrm{Re}\{\tilde K_{vv}(i\omega)\}$ for
the bistable potential with $\xi=1$ and for the monostable potential with $\xi=2\sqrt{2}/3$ is plotted in in Fig. \ref{fig:spect} for a selection of $S_{a*}$-values. 
 For both potentials, the spectra demonstrate an increased broadening and a tendency of downwards-in-frequency shift of the spectral weight. Despite their differences, these two potentials gave very similar performance in Fig. \ref{fig:pvs} and also display similar spectral shapes here. If we consider the curve for $S_{a*}=1\cdot 10^{-3}$ in Fig. \ref{fig:spect}, we see that the choice $\omega_0=0.5\omega_\mathrm{s}$ for the linear system discussed above 
lies within the spectrum of the nonlinear device and therefore is a fair case to compare to. 

Even though we only considered simple phenomenological potentials (\ref{eq:quarticasym}), the broadening and flattening of the spectrum and the better-than-linear power-characteristic within an $S_a$-range replicate experiments on a device
with an asymmetric monostable potential \cite{Nguyen2010}.

\section{Conclusion}

We have shown that when driven by white noise, harvesters with nonlinear stiffness do not have the fundamental performance advantage over linear ones  that one could have expected from their wider spectrum. This followed for efficient devices from considerations on input power and for general coupling from 
power bounds. Numerical examples were given for weak coupling. The findings do not preclude advantages of nonlinear-stiffness harvesters subject to vibrations significantly different from wide band noise, e.g. off-resonance, sufficiently band-limited vibrations. Implementation constraints may render a nonlinear stiffness unavoidable or a desired value of linear stiffness unattainable. We demonstrated advantages when linear stiffness was constrained. 

\begin{acknowledgments}
  I thank Prof. J.T. Scruggs for useful correspondence. This work was funded by The Research Council of Norway under grant no. 191282.
\end{acknowledgments}

%

\end{document}